# Controlling hybrid nonlinearities in transparent conducting oxides via two-colour excitation

M. Clerici,[1] N. Kinsey,[2,*] C. DeVault,[3] J. Kim,[2] E. G. Carnemolla,[4] L. Caspani,[4,5] A. Shaltout,[2,**] D. Faccio,[4] V. Shalaev,[2] A. Boltasseva,[2,+] and M. Ferrera[4,‡]

[1] School of Engineering, University of Glasgow, Glasgow, G12 8LT, UK

[2] School of Electrical and Computer Engineering and Birck Nanotechnology Center, Purdue University, West Lafayette, IN, 47907, USA

[3] Dept. of Physics & Astronomy and Birck Nanotechnology Center, Purdue University, West Lafayette, IN, 47907, USA

[4] Institute of Photonics and Quantum Sciences, Heriot-Watt University, SUPA, Edinburgh, Scotland, EH14 4AS, UK

[5] Institute of Photonics, Department of Physics, University of Strathclyde, Glasgow G1 1RD, UK

[*] Now with: Dep. of Electrical and Computer Engineering, Virginia Commonwealth University, Richmond, Virginia 23284, USA

[**] Now with: Geballe Laboratory for Advanced Materials, Stanford University, Stanford, California 94305, USA

[+,‡] Corresponding authors: aeb@purdue.edu; m.ferrera@hw.ac.uk.

## ABSTRACT

Nanophotonics and metamaterials have revolutionised the way we think about optical space $(\varepsilon, \mu)$, enabling us to engineer the refractive index almost at will, to confine light to the smallest of the volumes, and to manipulate optical signals with extremely small footprints and energy requirements. Significant efforts are now devoted to finding suitable materials and strategies for the dynamic control of the optical properties. Transparent conductive oxides exhibit large ultrafast nonlinearities under both interband and intraband excitations. Here, we show that combining these two effects in aluminium-doped zinc oxide via a two-colour laser field discloses new material functionalities. Owing to the independence of the two nonlinearities the ultrafast temporal dynamics of the material permittivity can be designed by acting on the amplitude and delay of the two fields. We demonstrate the potential applications of this novel degree of freedom by dynamically addressing the modulation bandwidth and optical spectral tuning of a probe optical pulse.

## INTRODUCTION

The continual success of electronics is largely due to the extreme miniaturization of devices and the ability to achieve exceptional functionality with a limited number of constituent materials. Currently, these areas are major weaknesses of photonics which hinder the primary advantage of increased operational bandwidth[1–3]. Specifically, the diffraction limit sets the minimum size of photonic components to about half their operational wavelength, while a plethora of photonic materials are needed to accomplish all the basic functionalities typically available in electronics. During the last decade, plasmonics and metamaterials have both gained great momentum because they provide a way to overcome the previously mentioned limitations. This is attained by coupling the electromagnetic radiation to the oscillating electronic plasma at the metal/dielectric interface.[4,5] However, the ability to squeeze optical modes well below the diffraction limit, and engineer the effective permittivity at will with structured materials is not free. By tightly confining light to metal layers, the propagation losses are increased which can be problematic for many applications.[6] To solve these relevant technological drawbacks, new material platforms have been recently investigated with the goal of removing metal constituents from both plasmonic technologies and artificial materials.[7,8]

Among alternative plasmonic materials, transparent conductive oxides (TCOs) are a unique class of wide-bandgap semiconductors. They can support extremely large doping levels ($\simeq 10^{21}$ cm$^{-3}$) with low effective electron masses ($\simeq 0.3 m_\mathrm{e}$; $m_\mathrm{e}$ = free-electron-mass), enabling a metallic response in the near-infrared (NIR) region and high transmission in the visible region.[8–12] Such materials have been employed as transparent electrodes in various applications,[13,14] but are recently receiving attention in nanophotonics, owing to their highly tunable static properties and potential for both electrical and optical control of the refractive index.[15–21] This versatility is extremely attractive as TCOs can serve multiple roles, for example as dynamic, plasmonic, and dielectric layers, which enable extreme flexibility to optimise structures for differing conditions, all with a single material. Furthermore, many TCOs are naturally suited to achieve the epsilon-near-zero (ENZ)[22] condition in the telecommunications band whereby the low refractive index, that is, less than unity, enables the potential for enhanced nonlinearities, super-coupling, and deeply sub-wavelength field confinement.[19,23–28] Aluminium-doped zinc oxide (AZO)[12] is one such material which combines all of these properties with reduced optical losses compared to other TCOs[8] and a low manufacturing cost due to its widely available elemental compounds. Consequently, oxygen deprived AZO (see Methods) is an attractive material for dynamic control. The optical injection of carriers into the conduction band (interband excitation in the ultraviolet –UV– spectral region, $\lambda_\mathrm{UV} = 325$ nm) enabled control of the reflection and transmission up to 40% and 35%, respectively, with sub-picosecond recovery time.[18] Intraband excitations with below-bandgap pulses (in the near-infrared –NIR– spectral region, $\lambda_\mathrm{NIR} = 787$ nm) resulted in up to 90% and 800% modulation of the reflection and transmission, respectively.

Here, we demonstrate that the optical properties of AZO can be dynamically addressed on a sub-picosecond time-scale by a clever combination of the interband and intraband nonlinear effects, driven by two different wavelength pump pulses ($\lambda_\mathrm{UV} = 262$ nm and $\lambda_\mathrm{NIR} = 787$ nm). We first show that the AZO complex refractive index at infrared wavelengths ($\lambda_\mathrm{p} = 1300$ nm) can be controlled by a two-colour pump pulse ($\lambda_\mathrm{UV} + \lambda_\mathrm{NIR}$). We then tune the delay between the two pumps to demonstrate the dynamic control of the probe modulation bandwidth, between 0.8 and 2 THz, and optical spectrum with a wavelength shift of $\pm 4$ nm.

## RESULTS

**Pump and probe experiments.** The pump-probe experiment is schematically shown in Figure 1a. See Methods for details on the experimental settings. We measure the probe pulse transmission (*T*) and reflection (*R*) from a 900 nm thick AZO film deposited on a 1 mm thick silica substrate. The probe pulse wavelength is set to $\lambda_\mathrm{p} = 1300$ nm, close to the zero-epsilon wavelength of the material. Both *T* and *R* are recorded as a function of the delay *Δτ* between the pump pulses and the probe. In Figure 1b we show the measured delay-dependent reduction of transmissivity as a result of interband carrier excitation, driven by an ultraviolet pulse ($\lambda_{UV} = 262$ nm) of $\simeq 65$ fs duration and fluence $F_{UV} \simeq 5$ mJ cm$^{-2}$. The recombination time ($\simeq 600$ fs) and modulation amplitude ($\simeq 45\%$) are consistent with previous measurements performed with a pump pulse at 325 nm.[18] In Figure 1c we show the measured delay-dependent increase of transmissivity due to intraband carrier excitation, driven by a 100 fs near-infrared pulse ($\lambda_{NIR} = 787$ nm) with fluence $F_{NIR} \simeq 16$ mJ cm$^{-2}$. This metal-like nonlinearity is the result of thermal smearing of free-electrons causing decreased absorption and reflection at the probe wavelength and exhibits an extremely fast decay rate of $\simeq 170$ fs. As shown by the dotted curves in Figs. 1b and 1c, the interband and intraband material responses can be successfully modelled using a Drude model coupled to the excess carrier population and the two-temperature model, respectively (see Methods).

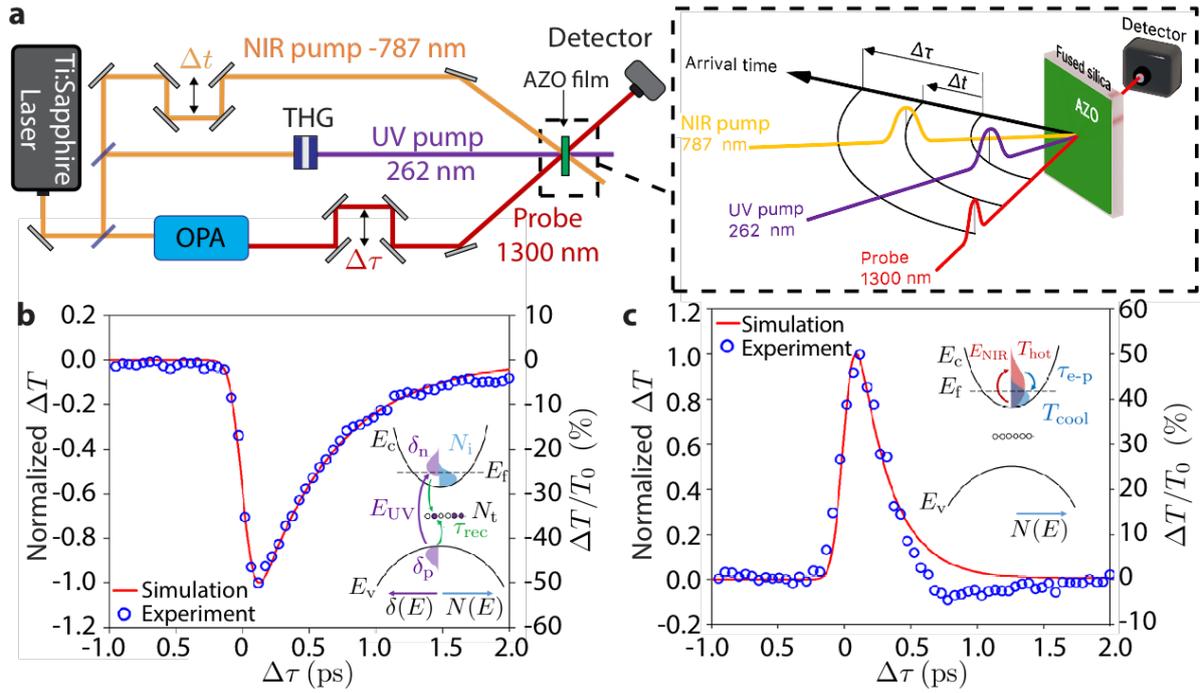

**Figure 1. Two-colour pump-probe experiment. a**, Schematic of the two-colour pump–probe experimental setup where the two pump wavelengths, 787 nm (near-infrared – NIR) and 262 nm (ultraviolet – UV), are illustrated along with the probe wavelength of 1300 nm. The delay between the two pump pulses is denoted $\Delta t$ while the delay between the probe and UV signal is denoted $\Delta\tau$. For intraband excitation using only the NIR pump, $\Delta\tau$ is defined as the delay between the probe and NIR pump pulse arrival time. The black arcs indicate the arrival time of the pulses. **b**, Change in transmission at 1300 nm versus the pump – probe delay $\Delta\tau$ under 262 nm excitation fitted with simulation. Inset illustrates the process diagram for interband excitation: UV light ($E_{UV}$) generates electron-hole pairs ($\delta_n, \delta_p$) above the Fermi level ($E_f$) in addition to the intrinsic concentration ($N_i$) which recombine through mid-gap trap states ($N_t, \tau_{rec}$). **c**, Change in transmission at 1300 nm versus the pump – probe delay $\Delta\tau$ under 787 nm excitation fitted with simulation. Inset illustrates the process diagram for intraband excitation: NIR light ($E_{NIR}$) raises the temperature of conduction band electrons ($T_{cool} \to T_{hot}$) which relax through scattering processes ($\tau_{e-p}$), heating the lattice. The change in transmission from the experiment was obtained through $\Delta T = T - T_0$, where $T_0$ is the transmission of the probe far from the pump pulse. For comparison with the theoretical results, the amplitude of the induced transmission change was normalized using $\Delta T_{\text{normalized}} = \Delta T/|\Delta T_{\text{max}}|$, where $|\Delta T_{\text{max}}|$ is the magnitude of the peak change in transmission. The same procedure was completed for the simulation results.

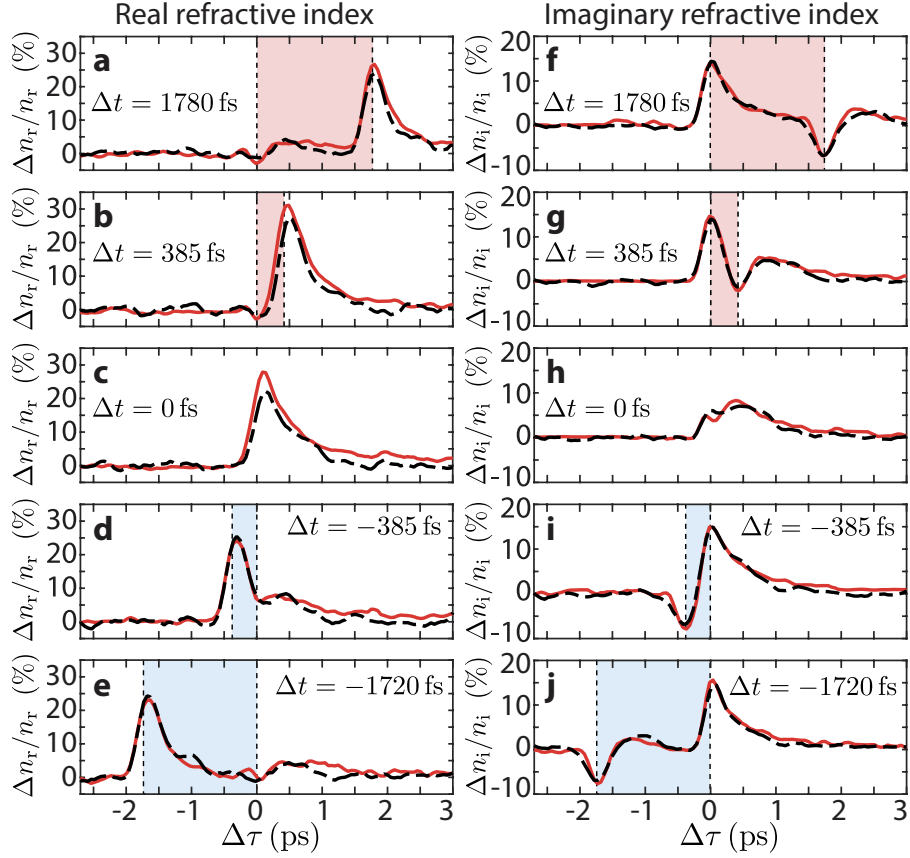

**Figure 2. Dynamic change in the optical response of AZO thin films triggered by two-colour excitation.** Percentage change of the real (**a-e**) and imaginary (**f-j**) part of the refractive index as a function of the delay $\Delta\tau$ between the ultraviolet (UV) pump and the infrared probe pulses. Multiple vertical plots are shown, for different delay $\Delta t$ between the UV and the near-infrared (NIR) pump pulses. Shaded areas indicate whether the UV pump, which is used as time reference, precedes (light blue) or follows (light red) the NIR pump. Overlapping the results obtained by simultaneous two-colour pumping (red curves) we plot the computed change in refractive index calculated by the algebraic summation of the results obtained from experiments with separate UV and NIR pump pulses (black dashed curves). The probe, the NIR, and the UV pump wavelengths were set to 1300 nm, 787 nm and 262 nm, respectively. The probe intensity was low: $I_\mathrm{p} \simeq 200$ MW cm$^{-2}$, while the pump fluences were $F_\mathrm{UV} = 5$ mJ cm$^{-2}$ and $F_\mathrm{NIR} = 14$ mJ cm$^{-2}$.

Remarkably, AZO shows simultaneously strong ultrafast interband and intraband nonlinearities with readily available wavelengths from a single laser source

The observed coexistence of both intraband and interband nonlinearities suggests the possibility of achieving new dynamic functionalities through their combined effect and if these two excitation regimes are independent, the corresponding effects can be algebraically combined. We therefore performed a thorough investigation of the AZO optical response under combined UV and NIR excitation as a function of the relative delay $\Delta t$ between the two pumps and $\Delta\tau$ between the pumps and the probe (the UV pump signal is used as the time reference). From the recorded probe reflection and transmission as a function of the two delays, $R_\mathrm{p}(\Delta t, \Delta\tau)$ and $T_p(\Delta t, \Delta\tau)$, respectively, (see Supplementary Figure 1), we retrieve the real and imaginary refractive index of the film using a transfer-matrix approach, as described elsewhere[29]. The extracted values are shown in Figure 2, where the time-dependent relative change in the real (Fig. 2a-e) and imaginary (Fig. 2f-j) refractive index of the AZO film at 1300 nm is shown, for five values of the NIR-UV pump pulse delay, between $\Delta t = -1.7$ ps and $\Delta t = 1.8$ ps. The results in Fig. 2 are achieved with optical pump fluences of $F_{\text{UV}} = 5$ mJ cm$^{-2}$ and $F_{\text{NIR}} = 14$ mJ cm$^{-2}$.

Our measurements demonstrate that the temporal dynamics of AZO film properties such as reflection and transmission, or equivalently the real and imaginary part of the refractive index, can be optically controlled via a two-colour excitation scheme. This is enabled by the independence of the two nonlinear processes responsible for the modulation of the material properties. This independence is demonstrated by the good match of the red and the black dashed curves in Fig. 2. The former are obtained from the measurements with simultaneous UV and NIR excitation while the latter are generated by the algebraic addition of the time-dependent refractive index changes induced by the UV and NIR pumps independently.

It is worth mentioning that all the experiments are performed in a condition of balanced excitation, meaning that the adopted fluences for UV and NIR pumping were set in such a way to produce similar alterations (in amplitude but not in sign) on the transmitted power. Operatively speaking, we first arbitrarily choose a fluence for the UV beam while the NIR pump fluence is set afterwards, to induce a change on the transmitted power of the same magnitude as that obtained with the UV pump. Such a change saturates for both UV ($\Delta T/T_0 \simeq 75$ % at $F_{\text{UV}} \simeq 15$ mJ cm$^{-2}$) and NIR ($\Delta T/T \simeq 100$ % at $F_{\text{NIR}} \simeq 60$ mJ cm$^{-2}$) pumping.

While cross-coupling between interband and intraband effects is negligible at the pump fluences used in these experiments, we observed that it becomes appreciable for higher fluences. To evaluate the impact of crosstalk we calculate the relative difference between the measured (me) and ideal (id) change in the real and imaginary refractive index $D_{\text{r,i}} = \left| n_{\text{r,i}}^{\text{me}} - n_{\text{r,i}}^{\text{id}} \right| / n_{\text{r,i}}^{\text{id}}$ for increasing pump fluences. In this case, "measured" refers to the refractive index with the simultaneous pumping scheme while "ideal" refers to the algebraic sum of the refractive indices retrieved with independent UV and NIR excitations. In the experimental conditions of Fig. 2 we estimate $D_{\text{r}} \leq 6$ % and $D_{\text{i}} \leq 3$ %, whereas at higher pump fluences $F_{\text{UV}} = 10$ mJ cm$^{-2}$ and $F_{\text{NIR}} = 24$ mJ cm$^{-2}$ we observe stronger crosstalk: $D_{\text{r}} \leq 22$ % and $D_{\text{i}} \leq 7$ % (see Supplementary Note 1 and Supplementary Figure 2).

**Modulation bandwidth and wavelength control.** The ability to optically control the AZO properties with nonlinear effects of similar amplitude yet opposite sign paves the way to intriguing applications. In Fig. 3 we show two new effects enabled by the two-colour AZO modulation. The first, in Figs. 3a and 3b, is the dynamic control of the optical modulation bandwidth of the AZO film, while the second, in Figs. 3c and 3d, is the dynamic control of the transmitted probe wavelength.

All-optical modulation of infrared radiation is relevant to the development of future telecommunications and data networks technologies[30–32]. An ultrafast, > 1 THz modulation bandwidth of AZO via interband pumping has been recently reported[18]. Figure 3a shows the change in $T$ of the AZO film pumped by both UV and NIR light, resolved as a function of the inter-pump delay $\Delta t$ and the pump-probe delay $\Delta\tau$. The effect of the UV pump is to reduce the transmission (the horizontal blue and purple band), while the effect of the NIR pump is to increase the transmission (the diagonal, green-to-red band). Performing the Fourier transform along the vertical direction provides the modulation bandwidth of the optically excited film as a function of the inter-pump delay $\Delta t$ and is shown in Fig. 3b. The blue and red dashed lines indicate the bandwidth (rms) obtained by only UV and only NIR pump, respectively. The faster dynamic of the intraband ($\simeq 1.7$ THz) compared to the interband ($\simeq 1.55$ THz) nonlinearity is clear. The proposed two-colour pump configuration remarkably allows one to modify the modulation bandwidth of the film via the delay of the two pump fields, as shown by the black curve in Fig. 3b. This enables the observation of a fast oscillation between a reduction ($\simeq 0.75$ THz at $\Delta t \simeq 0$) and an increase ($\simeq 2$ THz at $\Delta t \simeq 130$ fs) of the bandwidth, although with a four-fold reduction in the modulation depth compared to larger delays.

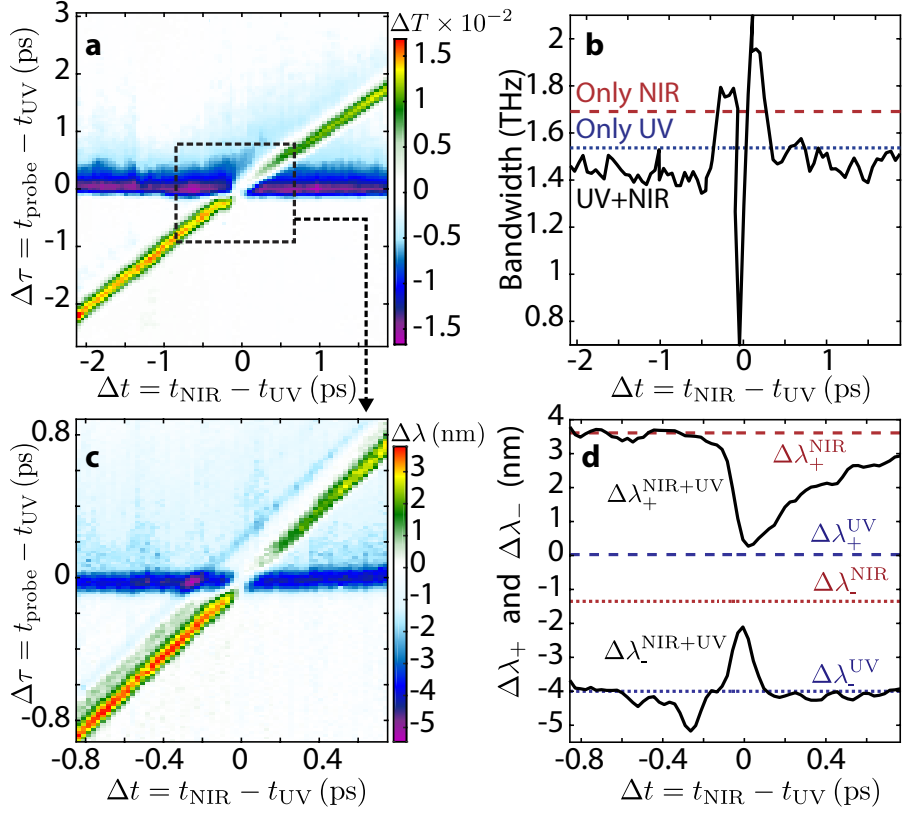

**Figure 3. Two-colour pumping effects. a**, Change in the probe pulse transmission as a function of the pump-probe delay ($\Delta\tau$) and the inter-pump delay ($\Delta t$). The UV pump decreases the transmission, whereas the NIR increases it. The combined effect produces a $\Delta t$-dependent modulation. The modulation bandwidth is evaluated by performing a Fourier transform along $\Delta\tau$, and is shown by the black curve in **b**. For delays close to $\Delta t = 0$ the modulation bandwidth can be decreased or increased by the two-colour combined effect. The blue and red dashed lines show the bandwidth of the UV-only and NIR-only driven modulation. **c**, Measured central wavelength shift of the transmitted probe pulse ($\simeq 15$ nm bandwidth) in a zoomed $\Delta t - \Delta\tau$ region (square box in **a**). The UV pump blue-shifts the probe wavelength, whereas the NIR pump does the opposite. At $\Delta t \simeq 0$ the opposite effects almost entirely cancel the wavelength shift. **d**, Summary of the findings in **c**, showing the maximum positive ($\Delta\lambda_+$) and negative ($\Delta\lambda_-$) wavelength shift for the NIR-only (red dashed/dotted), UV-only (blue dashed/dotted) and two-colour (solid black) AZO excitation.

Further, we show how the probe wavelength can be dynamically modified by a combination of the two UV and NIR pump pulses. In Fig. 3c we show the change in the central wavelength of the $\lambda_\mathrm{p} \simeq 1300$ nm probe pulse, recorded with an InGaAs spectrometer, as a function of both $\Delta t$ and $\Delta \tau$ and for pump fluences $F_\mathrm{UV} = 22 \text{ mJ cm}^{-2}$ and $F_\mathrm{NIR} = 42 \text{ mJ cm}^{-2}$. We note that the wavelength shift induced by the UV pump is negative, while the NIR pump gives both a positive and negative shift, depending on the delay with the probe. Figure 3d shows the maximum positive ($\Delta\lambda_+$) and negative ($\Delta\lambda_-$) frequency shift induced by the UV and NIR pumps alone (blue and red curves, respectively) and combined (black curves). Interestingly, the wavelength shift induced by the two independent excitation mechanisms can also be algebraically added. Therefore, for a specific choice of the pump fluences, the wavelength shift is almost cancelled when the two pump pulses are temporally overlapped. We note that, for higher pump fluences, wavelength shifts exceeding the pulse bandwidth can be achieved.

## *DISCUSSION*

AZO features both an interband and intraband transition at wavelengths that are different enough to separate the two effects, yet close enough to be addressed simultaneously by frequency conversion of standard laser sources. This is in contrast with most metals and semiconductors where the nonlinearities are either spectrally overlapped (metals), difficult to access (deep-UV), or spaced too far apart to enable their combined use (semiconductors). However, the observed properties are not unique to AZO, rather a common feature to the class of transparent conductive oxides. In fact, we observe similar behaviours, that is, the co-existence of interband and intraband nonlinearities with opposite optical effects, in commercially available indium tin oxide (ITO) films (see Supplementary Note 2 and Supplementary Figure 3). Although a thorough comparative analysis of the two TCOs goes beyond the purposes of this work, oxygen-deprived AZO shows stronger and faster temporal dynamics. Therefore, it is one of the best options for applications relying on the proposed hybrid nonlinearity at telecom wavelengths.

To conclude, we highlight that the large and dynamically addressable modulation of the AZO refractive index with terahertz bandwidths is a new photonic capability that may lead to interesting developments in signal processing, ultrafast optics, and optical metrology. For example, it may enable complex coding schemes such as code division multiple access[33], which is of high interest for optical communication. Similarly, engineering the recombination rate to match the thermal relaxation rate of the intraband excitation may lead to the implementation of an all-optical XOR gate via two-colour excitation. Finally, the observed ultrafast and controllable wavelength shift may lay the basis for a novel class of ultrafast and compact optical routers.

# *METHODS*

**Pump-probe experiment**. For the experiment we relied on a Ti:Sapphire laser (Amplitude Technologies), which delivered up to 10 mJ energy pulses at 787 nm central wavelength, with 100 fs pulse duration. A fraction of the laser power was routed to a commercial white light seeded optical parametric amplifier (Topas, Light Conversion Ltd), which produced the short ($<$ 120 fs) probe pulse at a wavelength tunable between 1100-2600 nm. For the experiments, we controlled the wavelength in the 1250-1350 nm range. The s-polarised probe beam was spatially filtered, reduced in energy with neutral density filters, and focused by a 250 mm focal length lens onto the AZO film, at a small angle to the normal of the film surface ($<$ 10 deg). The probe beam waist on the sample was measured using the knife-edge technique to be 65 μm. The probe intensity in the focus was found to be 200 MW cm$^{-2}$, determined by measuring the pulse energy and duration. The probe pulse delay from the pump pulses was set by a computer controlled linear translation stage (M-VP-25XA, Newport), equipped with a gold-coated hollow retroreflector (PLX Inc.).

The UV pulse (262 nm) was produced by pumping an in-line third-harmonic generation setup (Femtokit, Eksma Optics) with $\simeq$ 1 mJ, 787 nm pulses. The output 200 μJ, $\simeq$ 65 fs, 262 nm pulses were cleaned from the residual 787 nm and 393 nm radiation using four dichroic mirrors (HR @266 nm, HT@400 nm & @800 nm, Layertec GmbH). The s-polarised UV pulse was focused at normal incidence with a $f = 250$ mm focal length CaF$_2$ lens. The beam size measured with a knife-edge was 400 μm. The UV energy was controlled by acting on the half-wave plate at the input of the third-harmonic generation tool. The NIR pump was obtained by splitting a portion of the laser beam and delaying it from the UV pulse using a computer controlled linear translation stage (M-VP-25XA, Newport), equipped with a silver-coated hollow retroreflector (PLX Inc.). The s-polarised NIR pump was focused at normal incidence onto the AZO film. A dichroic mirror (HR @266 nm, HT @800 nm, Layertec GmbH) was employed for combining the two pump pulses. The beam waist of the NIR pump was measured to be 210 μm using the knife-edge technique. The pulse energy was controlled with a waveplate in front of a thin-film polariser (Altechna).

The UV, NIR and probe energies were measured with a thermopile detector (XPL12, Gentec-EO). The reflected and transmitted signals were recorded with amplified Germanium photodetectors (PDA50B-EC, Thorlabs).

**Sample fabrication and characterization**. The oxygen deprived aluminium-doped ZnO (AZO) films were deposited using pulsed laser deposition (PVD Products Inc.) with a KrF excimer laser (Lambda Physik GmbH) operating at a wavelength of 248 nm for source material ablation. A 2wt % doped AZO target was purchased from the Kurt J. Lesker Corp. with a purity of 99.99 % or higher. The energy density of the laser beam at the target surface was maintained at 1.5 J cm$^{-2}$ and the deposition temperature was 75°C. We maintained the oxygen pressure under 0.01 mTorr to achieve additional free carriers from the oxygen vacancies. The prepared thin films were characterised by spectroscopic ellipsometry (J. A. Woollam Co. Inc.) in the spectral region from 300 - 2500 nm. The dielectric function of the AZO was retrieved by fitting a Drude (when $\omega_o = 0$) and Lorentz oscillator model, Eq. (1), to the ellipsometry data. The optical properties at 262 nm were estimated from a spline extrapolation of the measured properties combined with bounds provided by data from similar films.[34] To probe the electrical properties of thin films such as mobility and carrier concentration, we carried out the Hall measurement (MMR Technologies) at room temperature.

$$\varepsilon = \varepsilon_\infty + \sum \frac{\omega_p^2}{(\omega_o^2 - \omega^2) - i\omega\Gamma} \qquad (1)$$

**Interband dynamics and modelling.** The interband modulation, $E_{h\nu} > E_g$, and relaxation of the AZO film were first observed individually using only the 262 nm UV pump and probe. The sub-picosecond recombination time is indicative of Shockley-Read-Hall recombination processes, which result in a reduced recombination time according to $\tau_{rec} \propto 1/N_t \sigma v_{th}$, where $N_t$ is the trap density, $\sigma$ is the capture cross section, and $v_{th}$ is the thermal velocity of carriers.[35] The interband dynamics were modelled using a 2D spatial and temporal discretization (see Supplementary Discussion for details).

The change in the optical properties was then determined using the transfer matrix method for the graded index profile whereby a matrix was calculated for each layer $\delta z$ and multiplied to determine an effective transfer matrix at each time step assuming an infinite substrate of fused silica. The amplitude of the change was then normalized and the recombination rate was fitted to the experimental data. Subsequently, the recombination time of the film was estimated to be $\simeq 600$ fs.

**Intraband dynamics and modelling**. The intraband dynamics, $E_{h\nu} < E_g$ of the AZO film were also observed individually using only the 787 nm excitation and probe. At 787 nm the AZO is a lossy dielectric, but the excitation is still far from the band-edge ($\lambda \simeq 320$ nm). Subsequently, the absorption in this regime is dominated by the residual Drude loss, i.e., free carriers in the conduction band. This excitation results in a non-equilibrium hot electron population which relaxes through various scattering processes ($\tau_{e-p}$), heating the lattice. Consequently, the intraband dynamics of the AZO film were modelled using the two-temperature model, whereby the change in the electron temperature and lattice temperature are captured as a function of time for the material (see Supplementary Discussion for details). Following, the resulting change in the optical properties was modelled using an effective thermally dependent complex index, $n_{th}$, such that $\Delta n_{AZO} = (\Delta T_e + \Delta T_l) n_{th}$. The transfer matrix method was used to determine the change in the optical properties of the graded index material. The reflection and transmission of the sample were normalized, and the rate was fitted to extract the electron phonon-coupling coefficient and found to be $G \simeq 1.4 \; 10^{16} \; W \; K^{-1} \; m^{-3}$. After the normalisation procedure, only the sign of the complex effective thermal index is relevant, and it was found that the extinction coefficient decreases while the index increased, i.e. $n_{th} > 0$ and $k_{th} < 0$, matching the experimental results. However, unlike other materials, we did not observe a long-lasting thermal offset arising from heat dissipation and removal from the lattice. This is likely due to the significant disparity between the electron and lattice heat capacities of AZO (see Supplementary Discussion). Thus, for a small application of energy the electron temperature increases significantly, giving rise to a large nonlinear response and an observable change in the reflection/transmission, although this energy does not result in a substantial increase in the lattice temperature.

**Data availability.** All relevant data present in this publication can be accessed at: http://dx.doi.org/10.17861/8f45636c-0560-427b-992e-87ba6d9090ab.

**Acknowledgements.** This work was supported by the NSF MRSEC Grant DMR-1120923, AFOSR Grant FA9550-14-1-0138, and AFOSR Grant FA9550-14-1-0389. D.F. acknowledges financial support from the European Research Council under the European Union's Seventh Framework Programme (FP/2007-2013)/ERC GA 306559 and EPSRC (UK, grant EP/M009122/1). M.C. acknowledges support from EPSRC (Grants No. EP/P009697/1 and EP/P51133X/1) and from BLM s.p.a.. L.C. acknowledges the support from the People Programme (Marie Curie Actions) of the European Union's FP7 Programme under REA Grant Agreements No. 627478 (THREEPLE). M. F. acknowledges sponsorship from EPSRC (UK, grant EP/P019994/1). The authors thank Prof. Arrigo Calzolari, CNR Nano Modena Italy, for helpful discussions during the development of numerical models

**Author contributions.** M.C., E.G.C., L.C., D.F., M.F. contributed to the optical characterization and measurements. N.K., C.D., J.K., A.S., V.S., A.B. contributed to the material development and modelling. All the authors contributed to the writing of the manuscript